# Performance Constrained Distributed Event-triggered Consensus in Multi-agent Systems

Amir Amini, *Student Member, IEEE*, Zahra Zeinali, *Student Member, IEEE*, Amir Asif, *Senior Member, IEEE*, Arash Mohammadi, *Senior Member, IEEE*

***Abstract*— The paper proposes a distributed event-triggered consensus approach for linear multi-agent systems with guarantees over rate of convergence, resilience to control gain uncertainties, and Pareto optimality of design parameters, namely, the event-triggering threshold (ET) and control gain. The event-triggered consensus problem is first converted to stability problem of an equivalent system. The Lyapunov stability theorem is then used to incorporate the performance constraints with the event-triggered consensus. Using an approximated linear scalarization method, the ET and the control gain are designed simultaneously by solving a convex constrained optimization problem. Followed by some preliminary steps, the optimization can be performed locally, i.e., no global information is required. The effectiveness of the proposed approach is studied through simulations for an experimental multi-agent system.**

## I. Introduction

Among cooperative behaviours in multi-agent systems, consensus [1], [2] has attracted overwhelming attention due to its widespread application in several engineering domains of significant importance including, but not limited to, control of autonomous vehicles such as unmanned aerial vehicles (UAV) [3] and autonomous underwater vehicles (AUV) [4]. To preserve the valuable on-board energy resources allocated to each agent, event-triggered strategies have been recently developed to decrease data transmission and/or control input updates in the consensus process. Numerous event-triggered schemes [5] have been recently implemented for a wide variety of consensus problems from different perspectives. Despite introducing several advantageous properties, existing event-triggered consensus strategies face at least one of the following three shortcomings:

Firstly, it is a common assumption in the context of event-triggered consensus that the control gains are implementable with infinite precision. Beside the round-off error, it is widely known that the control parameters are often subject to uncertainties due to physical constraints [6]. The uncertainties in realization of the control gains can cause considerable performance deterioration. In this regard, *resilient* (*non-fragile*) control design techniques, which take into account the uncertainties of the controller realization, are considered in various control problems [6], [7]. We note that since non-real-time and limited neighbouring information is used in event-triggered schemes, the closed-loop system is more vulnerable to such uncertainties. To the best of our knowledge, however, distributed implementation of non-fragile control design has not yet been investigated for event-triggered consensus. The paper addresses this gap.

Secondly, the implementation of an event-triggered scheme often requires a control gain, used by the control law, and an event-triggering threshold (ET) used by the event detectors. Due to interactive operation of the control law and event-detectors that mutually affect consensus performance, design of the ET and control gain should be accomplished through a *unified* framework. As the second limitation in most strategies, control gain is either assumed a priori [8], [9] or designed as the first step based on different interpretations of Hurwitz stability. Then, as the second step, an operating region is developed for the ET [10]–[12] which depends on the first step (i.e., the control gain design). In an impartial approach, however, the ET and control gain should be designed simultaneously w.r.t an objective function considering both parameters. As stated in [5], it is a challenging issue to design both the ET and control gain in a unified framework. The non-existence of an efficient approach to co-design these parameters through a distributed single step optimization is the second motivation of this work.

Third, an important issue in distributed networked control systems is the *convergence rate* of the proposed implementation to the desired objective [13]. In this regard, a flexible rate or a finite-time convergence is of great importance. Most existing event-triggered consensus approaches, however, ensure only an asymptotic convergence rate which may be conservative in some cases. We note that ensuring a minimum rate of consensus in general multi-agent systems is a non-trivial task when the transmission scheme is event-triggered.

In summary, the key contributions of the paper are threefold: (i) The proposed approach is resilient to control gain uncertainties and guarantees a desired exponential rate of consensus. (ii) Consensus parameters, i.e., the ET and control gain, are computed simultaneously through a convex constrained optimization problem. An approximated linear scalarization objective function is proposed to optimize consensus parameters. (iii) In contrast to our previous work [14], the optimization stage in this paper only requires some parameters that can be estimated locally. Therefore, no fusion center is needed to perform the optimization. Moreover, complexity of the proposed optimization is irrespective of the network size which vastly improves scalability of the approach.

## II. Preliminaries and Problem Statement

Throughout the paper, a bold alphabet indicates either a matrix or a vector, and normal letters stand for scalars. In Table I, we list necessary notation used in the paper.

Consider a general linear time-invariant multi-agent system described by the following dynamics

$$\dot{\boldsymbol{x}}_i(t) = \boldsymbol{A}\boldsymbol{x}_i(t) + \boldsymbol{B}\boldsymbol{u}_i(t), \quad (1 \leq i \leq N), \tag{1}$$

where $\boldsymbol{x}_i(t) \in \mathbb{R}^n$ and $\boldsymbol{u}_i(t) \in \mathbb{R}^m$ are, respectively, the state and control input. The pair $(\boldsymbol{A}, \boldsymbol{B})$ is controllable. The consensus problem is said to be solved iff $\lim_{t\to\infty}\|\boldsymbol{x}_i(t)-\boldsymbol{x}_j(t)\| = 0$, $(1 \leq i, j \leq N)$ [10]. Through an undirected connected network, the agents share their state values within their neighbourhood to achieve consensus. To reduce data transfer, an event-detector is incorporated with each agent. The event-detector at node $i$ monitors an 'event-triggering condition' to determine whether or not to transmit the real-time state value $\boldsymbol{x}_i(t)$. If the event detector detects an event at time instant $t^i_{k^{[i]}}$ (superscript $i$ indicates node $i$, and subscript $k^{[i]} = 0, 1, \cdots$ denotes the sequence of events for agent $i$), then node $i$ transmits $\boldsymbol{x}_i(t^i_{k^{[i]}})$ to its neighbours. Upon receiving $\boldsymbol{x}_i(t^i_{k^{[i]}})$, agent $j$, $(j \in \mathcal{N}_j)$, updates its database. This state value, i.e., $\boldsymbol{x}_i(t^i_{k^{[i]}})$, is used at node $j$ until the next event is triggered from node $i$. Let $\hat{\boldsymbol{x}}_i(t) \triangleq \boldsymbol{x}_i(t^i_{k^{[i]}})$, $t \in [\,t^i_{k^{[i]}},\, t^i_{k^{[i]}+1}\,)$. The disagreement for agent $i$, $(1 \leq i \leq N)$, is defined as follows

$$\mathbb{X}_i(t) = \sum_{j\in\mathcal{N}_i} a_{i,j}\left(\boldsymbol{\Lambda}_i(t)\hat{\boldsymbol{x}}_i(t) - \boldsymbol{\Lambda}_j(t)\hat{\boldsymbol{x}}_j(t)\right), \tag{2}$$

where $\boldsymbol{\Lambda}_i(t) = e^{\boldsymbol{A}(t - t^i_{k^{[i]}})}$. Note that $\boldsymbol{\Lambda}_i(t)\hat{\boldsymbol{x}}_i(t)$ is an open-loop estimate of $\boldsymbol{x}_i(t)$ in $t \in [\,t^i_{k^{[i]}},\, t^i_{k^{[i]}+1}\,)$ [15]. Allowing all agents to transmit their initial state values $\boldsymbol{x}_i(0)$ to the neighbours, the norm of $\mathbb{X}_i(t)$ is used in the local event detectors to determine the next event. Given $t^i_{k^{[i]}}$, the next event for agent $i$, $(1 \leq i \leq N)$, is triggered based on the following event-triggering condition

$$t^i_{k^{[i]}+1} = \inf\{\, t > t^i_{k^{[i]}} : \|\boldsymbol{e}_i(t)\| - \phi\, \|\mathbb{X}_i(t)\| \geq 0\}. \tag{3}$$

Scalar $\phi > 0$ is the ET to be designed. Vector $\boldsymbol{e}_i(t) = \boldsymbol{\Lambda}_i(t)\hat{\boldsymbol{x}}_i(t) - \boldsymbol{x}_i(t)$, $(1 \leq i \leq N)$, denotes the state error. The following distributed control law [15] is proposed as the forcing term for (1)

$$\boldsymbol{u}_i(t) = (\boldsymbol{K} + \boldsymbol{\Delta}_{K\,i}(t))\,\mathbb{X}_i(t), \quad (1 \leq i \leq N). \tag{4}$$

TABLE I: Mathematic and graph notation.

| Notation | Definition | Notation | Definition |
|---|---|---|---|
| $\|.\|$ | Frobenius norm | $(.)^\dagger$ | Pseudo inverse |
| $\boldsymbol{A}>0$ | Matrix $\boldsymbol{A}$ is symmetric positive definite | $\mathbf{1}_n$ | Column vector of order $n$ with all entries equal to 1 |
| $\boldsymbol{I}$ | Identity matrix | $\mathbf{0}$ | Zero matrix |
| $\otimes$ | Kronecker product | $\mathcal{A}=\{a_{i,j}\}$ | Weighted adjacency matrix |
| $\boldsymbol{L}$ | Laplacian matrix | $\mathcal{N}_i$ | Set of neighbours for node $i$ |

Matrix $\boldsymbol{K} \in \mathbb{R}^{m\times n}$ is the control gain to be designed. The time-varying matrix $\boldsymbol{\Delta}_{K\,i}(t)$ represents uncertainties in the control gain which often happen due to implementation constraints, actuator faults, system modeling and digitalization errors [6].

**Assumption 1.** The uncertainty $\boldsymbol{\Delta}_{K\,i}(t)$ satisfies $\|\boldsymbol{\Delta}_{K\,i}(t)\| \leq \delta_i$, $(1 \leq i \leq N)$, for $t > 0$, where $\delta_i$ is a local predetermined desired resilience level for $\boldsymbol{K}$.

**Design Objectives:** It is evident from (3) that an extremely small $\phi$ leads to almost constant transmission, whereas an extremely large $\phi$ may delay achievement of consensus (due to lack of sufficient amount of transmission) or even endanger the closed-loop stability. Therefore, the value of $\phi$ which does not prevent a desired rate of consensus and, at the same time, efficiently reduces the transmission load is of great importance. On the other hand, although a large norm for $\boldsymbol{K}$ can often accelerate the consensus convergence, it forces undesirable large control inputs. It can also be shown (Remark 2) that larger control gains result it more dense events. Due to such interaction between the design parameters, we propose a *unified* optimization framework to simultaneously optimize $\phi$ and $\boldsymbol{K}$ w.r.t a proposed objective function. The design objectives can, therefore, be considered in threefold: (i) Upper-bound the rate of consensus convergence by an exponentially decreasing term; (ii) Guarantee resilience to some level of uncertainty in the control gain, and; (iii) Limit the norm of $\boldsymbol{K}$ (to prevent large control inputs and reduce the number of events) and enlarge $\phi$ (to reduce the number of events).

## III. problem formulation

Let $\boldsymbol{x}(t) = [\boldsymbol{x}^T_1(t), \ldots, \boldsymbol{x}^T_N(t)]^T$, $\hat{\boldsymbol{x}}(t) = [\hat{\boldsymbol{x}}^T_1(t), \ldots, \hat{\boldsymbol{x}}^T_N(t)]^T$, and $\boldsymbol{e}(t) = [\,\boldsymbol{e}^T_1(t)\,,\,\ldots\,,\,\boldsymbol{e}^T_N(t)\,]^T$. Accordingly, $\boldsymbol{e}(t) = \boldsymbol{\Lambda}(t)\hat{\boldsymbol{x}}(t) - \boldsymbol{x}(t)$, with $\boldsymbol{\Lambda}(t) = \mathrm{diag}(\boldsymbol{\Lambda}_1(t), \ldots, \boldsymbol{\Lambda}_N(t))$. Below is the closed-loop system from (1) and (4)

$$\dot{\boldsymbol{x}}(t) = (\boldsymbol{A}_{[N]} + \boldsymbol{B}_{[N]}\bar{\boldsymbol{K}}\boldsymbol{L}_{[n]})\,\boldsymbol{x}(t) + \boldsymbol{B}_{[N]}\bar{\boldsymbol{K}}\boldsymbol{L}_{[n]}\,\boldsymbol{e}(t), \tag{5}$$

where $\boldsymbol{A}_{[N]} = \boldsymbol{I}_N \otimes \boldsymbol{A}$, $\boldsymbol{B}_{[N]} = \boldsymbol{I}_N \otimes \boldsymbol{B}$, $\bar{\boldsymbol{K}} = \boldsymbol{K}_{[N]} + \boldsymbol{\Delta}_K$, and $\boldsymbol{L}_{[n]} = \boldsymbol{L} \otimes \boldsymbol{I}_n$, with $\boldsymbol{K}_{[N]} = \boldsymbol{I}_N \otimes \boldsymbol{K}$, and $\boldsymbol{\Delta}_K = \mathrm{diag}\,(\,\boldsymbol{\Delta}_{K\,1}(t)\,,\,\ldots\,,\,\boldsymbol{\Delta}_{K\,N}(t)\,)$. From Assumption 1, it holds that $\|\boldsymbol{\Delta}_K\| \leq \delta$ where $\delta = \sqrt{N}\max_{1\leq i\leq N}\delta_i$.

***A. Consensus to Stability***: A widely used approach to solve consensus is to convert the original consensus problem to stability problem of an equivalent system [11]. In this regard, the following state transformation is used to convert the consensus problem into stability problem

$$\boldsymbol{x}_{[r]}(t) = \hat{\boldsymbol{L}}_{[n]}\boldsymbol{x}(t). \tag{6}$$

where $\hat{\boldsymbol{L}}_{[n]} = \hat{\boldsymbol{L}} \otimes \boldsymbol{I}_n$, and $\hat{\boldsymbol{L}} \in \mathbb{R}^{(N-1)\times N}$ is obtained by removing one arbitrary row from $\boldsymbol{L}$. According to [14, Lemma 1], if the system expressed by (6) is stable then consensus is achieved in (5). Using (6), we convert (1) to

the following system

$$\dot{\boldsymbol{x}}_{[r]}(t) = \Big(\boldsymbol{A}_{[\hat{N}]} + \boldsymbol{B}_{[\hat{N}]}\boldsymbol{K}_{[\hat{N}]}\mathbb{L}_{[n]} + \hat{\boldsymbol{L}}_{[n]}\boldsymbol{B}_{[N]}\boldsymbol{\Delta}_K\boldsymbol{\mathcal{L}}_{[n]}\Big)\,\boldsymbol{x}_{[r]}(t) + \Big(\hat{\boldsymbol{L}}_{[n]}\boldsymbol{B}_{[N]}\boldsymbol{K}_{[N]} + \hat{\boldsymbol{L}}_{[n]}\boldsymbol{B}_{[N]}\boldsymbol{\Delta}_K\Big)\,\boldsymbol{L}_{[n]}\,\boldsymbol{e}(t), \tag{7}$$

where $\boldsymbol{A}_{[\hat{N}]} = \boldsymbol{I}_{N-1}\otimes\boldsymbol{A}$, $\boldsymbol{B}_{[\hat{N}]} = \boldsymbol{I}_{N-1}\otimes\boldsymbol{B}$, $\boldsymbol{K}_{[\hat{N}]} = \boldsymbol{I}_{N-1}\otimes\boldsymbol{K}$, $\mathbb{L}_{[n]} = \mathbb{L}\otimes\boldsymbol{I}_n$, and $\boldsymbol{\mathcal{L}}_{[n]} = \boldsymbol{\mathcal{L}}\otimes\boldsymbol{I}_n$, with $\mathbb{L} = \hat{\boldsymbol{L}}\boldsymbol{L}\hat{\boldsymbol{L}}^{\dagger}$, and $\boldsymbol{\mathcal{L}} = \boldsymbol{L}\hat{\boldsymbol{L}}^{\dagger}$. Without loss of generality and for brevity, we remove row $N$ from $\boldsymbol{L}$ to obtain $\hat{\boldsymbol{L}}$. The exponential stability is defined in what follows.

**Definition 1.** *Exponential Stability* [16]: Given a convergence rate $\zeta > 0$, system (7) is said to be $\zeta$-exponentially stable if there exists scalar $\eta > 0$ such that $\boldsymbol{x}_{[r]}(t)$ satisfies $\|\,\boldsymbol{x}_{[r]}(t)\,\| \leq \eta\, e^{-\zeta t}\|\,\boldsymbol{x}_{[r]}(0)\,\|$, $(t \geq 0)$, for any $\boldsymbol{x}_{[r]}(0)$.

According to consensus definition and transformation (6), the exponential stability for system (7) is equivalent to exponential consensus for (5).

To incorporate the event-triggering condition (3) with system (7), one should express (3) with respect to $\boldsymbol{x}_{[r]}(t)$ and $\boldsymbol{e}(t)$. Between two consecutive events for agent $i$, it holds that $\|\boldsymbol{e}_i(t)\| \leq \phi\|\mathbb{X}_i(t)\|$. In a collective fashion, the following element-wise inequality is derived from (3)

$$\boldsymbol{e}^{[\mathrm{nr}]} \leq \phi\,\mathbb{X}^{[\mathrm{nr}]}, \tag{8}$$

where $\boldsymbol{e}^{[\mathrm{nr}]} = [\,\|\boldsymbol{e}_1(t)\|, \ldots, \|\boldsymbol{e}_N(t)\|\,]^T$, $\mathbb{X}^{[\mathrm{nr}]} = [\,\|\mathbb{X}_1(t)\|, \ldots, \|\mathbb{X}_N(t)\|\,]^T$. The following Lemma is proposed to convert (3) in the desired form.

**Lemma 1.** Let $\boldsymbol{M} = [\,\boldsymbol{I}_{N-1},\ \ \boldsymbol{1}_N\,]^T$. If a certain value for $\phi$ satisfies the following entry-wise inequality

$$\boldsymbol{e}^{[\mathrm{nr}]} \leq \phi\,\boldsymbol{M}\,\hat{\mathbb{X}}^{[\mathrm{nr}]}, \tag{9}$$

with $\hat{\mathbb{X}}^{[\mathrm{nr}]} = [\|\mathbb{X}_1(t)\|, \ldots, \|\mathbb{X}_{N-1}(t)\|]^T$, it also satisfies (8).

*Proof.* Since $\mathbb{X}(t)$ is formed by the row space of $\boldsymbol{L}$, a specific element in $\mathbb{X}(t)$, e.g., $\mathbb{X}_N(t)$, is linearly dependent on other elements of $\mathbb{X}(t)$. More precisely, $\mathbb{X}_N(t) = \alpha_1\mathbb{X}_1(t) + \cdots + \alpha_{N-1}\mathbb{X}_{N-1}(t)$. For a symmetric connected network, it can be shown that $\alpha_j = -1$, $(1 \leq j \leq N-1)$. As a result, it holds that $\|\mathbb{X}_N(t)\| \leq \|\mathbb{X}_1(t)\| + \cdots + \|\mathbb{X}_{N-1}(t)\|$. Accordingly, the following inequality is derived from (8)

$$\boldsymbol{e}^{[\mathrm{nr}]} \leq \phi\,\mathbb{X}^{[\mathrm{nr}]} \leq \phi\,\boldsymbol{M}\,\hat{\mathbb{X}}^{[\mathrm{nr}]}. \tag{10}$$

Therefore, if (9) is satisfied for a specific $\phi$, inequality (8) is also satisfied using the same value of $\phi$. □

**Lemma 2.** Based on definition of $\mathbb{L}$, $\boldsymbol{\mathcal{L}}$, and $\hat{\boldsymbol{L}}$, these inequalities hold: i) $\lambda_{\min}(\mathbb{L}) = \lambda_2$, ii) $\lambda_{\max}(\hat{\boldsymbol{L}}) \leq \lambda_N$, iii) $\lambda_{\max}(\boldsymbol{\mathcal{L}}) \leq \lambda_N$, where $\lambda_2$ is the second smallest and $\lambda_N$ is the largest eigenvalue of $\boldsymbol{L}$.

To save on space, we omit proof of Lemma 2. We remove the time argument from vectors $\boldsymbol{x}_{[r]}(t)$ and $\boldsymbol{e}(t)$ to improve readability.

***B. Parameter Design***: Using the following theorem, we simultaneously compute the ET $\phi$ and control gain $\boldsymbol{K}$.

**Theorem 1.** Given desired $\zeta$ and $\delta$, the ET $\phi$ and control gain $\boldsymbol{K}$ are computed from the following equations

$$\phi = \sqrt{\tau_4{}^{-1}\gamma^{-1}}, \qquad \boldsymbol{K} = \boldsymbol{B}^{\dagger}\boldsymbol{\mathcal{P}}^{-1}\boldsymbol{\Omega}. \tag{11}$$

The validity of (11) is conditioned on the existence of symmetric negative definite matrix $\boldsymbol{\Omega} \in \mathbb{R}^{n\times n}$, symmetric positive definite matrix $\boldsymbol{\mathcal{P}} \in \mathbb{R}^{n\times n}$, positive scalars $\gamma$, $\tau_c$, $(1 \leq c \leq 4)$, $\epsilon_j$, $(1 \leq j \leq 3)$, and $\theta_c$, $(1 \leq c \leq 4)$, satisfying the following minimization problem

$$\min_{\boldsymbol{\Omega},\boldsymbol{\mathcal{P}},\gamma,\tau_c,\epsilon_j,\theta_c} \quad \mathcal{F} = \theta_1 + \theta_2 + \theta_3 + \theta_4, \tag{12}$$

S.t:

$$\boldsymbol{\Psi} = \begin{bmatrix} \boldsymbol{\Psi}_{11} & \boldsymbol{0} & \boldsymbol{0} & \boldsymbol{0} & \boldsymbol{I} & \|\boldsymbol{M}_{[n]}\|\boldsymbol{I} & \boldsymbol{0} & \boldsymbol{\Omega} & \boldsymbol{0} \\ * & \boldsymbol{\Psi}_{22} & \boldsymbol{0} & \boldsymbol{0} & \boldsymbol{0} & \boldsymbol{0} & \boldsymbol{0} & \boldsymbol{0} & \boldsymbol{0} \\ * & * & -\tau_1\boldsymbol{I} & \boldsymbol{0} & \boldsymbol{0} & \boldsymbol{0} & \boldsymbol{\mathcal{P}} & \boldsymbol{0} & \boldsymbol{0} \\ * & * & * & \boldsymbol{\Psi}_{44} & \boldsymbol{0} & \boldsymbol{0} & \boldsymbol{0} & \boldsymbol{0} & \boldsymbol{0} \\ * & * & * & * & -\tau_3\boldsymbol{I} & \boldsymbol{0} & \boldsymbol{\mathcal{P}} & \boldsymbol{0} & \boldsymbol{0} \\ * & * & * & * & * & -\gamma\boldsymbol{I} & \boldsymbol{0} & \boldsymbol{0} & \|\boldsymbol{M}_{[n]}\|\boldsymbol{I} \\ * & * & * & * & * & * & -\epsilon_1\boldsymbol{I} & \boldsymbol{0} & \boldsymbol{0} \\ * & * & * & * & * & * & * & -\epsilon_2\boldsymbol{I} & \boldsymbol{0} \\ * & * & * & * & * & * & * & * & -\epsilon_3\boldsymbol{I} \end{bmatrix} < 0, \tag{13}$$

$$\begin{bmatrix} \theta_1\boldsymbol{I} & \boldsymbol{I} \\ * & \boldsymbol{\mathcal{P}} \end{bmatrix} > 0, \qquad \begin{bmatrix} -\theta_2\boldsymbol{I} & \boldsymbol{\Omega}^T \\ * & -\boldsymbol{I} \end{bmatrix} < 0, \tag{14}$$

$$\tau_4 - \theta_3 < 0, \qquad \gamma - \theta_4 < 0, \tag{15}$$

where $\boldsymbol{\Psi}_{11} = \boldsymbol{A}^T\boldsymbol{\mathcal{P}} + \boldsymbol{\mathcal{P}}\boldsymbol{A} + 2\lambda_2\boldsymbol{\Omega} + \tau_1\delta^2\lambda_N^2\boldsymbol{I} + 2\zeta\boldsymbol{\mathcal{P}} + \epsilon_1\lambda_N^2\boldsymbol{B}\boldsymbol{B}^T$, $\boldsymbol{\Psi}_{22} = (-\tau_4 + \epsilon_3\lambda_N^2 + \tau_3\lambda_N^2\delta^2 + \tau_2\lambda_N^2)\boldsymbol{I}$, and $\boldsymbol{\Psi}_{44} = (-\tau_2 + \epsilon_2\lambda_N^2)\boldsymbol{I}$. Using parameters (11), consensus is achieved at the desired $\zeta-$exponential rate with $\eta = \lambda_{\max}(\boldsymbol{\mathcal{P}})/\lambda_{\min}(\boldsymbol{\mathcal{P}})$. The following inequalities are guaranteed for the minimized objective function $\mathcal{F}$

$$\boldsymbol{K}\boldsymbol{K}^T \leq \theta_1^2\,\theta_2\,\boldsymbol{B}^{\dagger}\,\boldsymbol{B}^{\dagger^T}, \qquad \phi \geq (\theta_3\,\theta_4)^{\frac{-1}{2}}. \tag{16}$$

*Proof.* To develop the $\zeta$-exponential stability conditions for (7), we consider the following inequality

$$\dot{V}(t) + 2\zeta V(t) < 0, \tag{17}$$

where $V(t) = \boldsymbol{x}_{[r]}^T\boldsymbol{\mathcal{P}}_{[\hat{N}]}\,\boldsymbol{x}_{[r]}$ with $\boldsymbol{\mathcal{P}}_{[\hat{N}]} = \boldsymbol{I}_{N-1}\otimes\boldsymbol{\mathcal{P}}$. From (17), it holds that

$$\lambda_{\min}(\boldsymbol{\mathcal{P}})\|\boldsymbol{x}_{[r]}\|^2 \leq V(t) < V(0)\,e^{-2\zeta t} \leq \lambda_{\max}(\boldsymbol{\mathcal{P}})\,e^{-2\zeta t}\|\boldsymbol{x}_{[r]}(0)\|^2.$$

The above inequalities lead to exponential stability, with the previously defined $\eta$. Therefore, inequality (17) is the sufficient constraint to ensure $\zeta$-exponential stability for (7). Let $\boldsymbol{\alpha} = [\,\boldsymbol{x}_{[r]}^T, \boldsymbol{e}^T, \boldsymbol{\sigma}_1^T, \boldsymbol{\sigma}_2^T, \boldsymbol{\sigma}_3^T\,]^T$, with $\boldsymbol{\sigma}_1 = \boldsymbol{\Delta}_K\boldsymbol{\mathcal{L}}_{[n]}\,\boldsymbol{x}_{[r]}$, $\boldsymbol{\sigma}_2 = \boldsymbol{L}_{[n]}\,\boldsymbol{e}$, and $\boldsymbol{\sigma}_3 = \boldsymbol{\Delta}_K\boldsymbol{L}_{[n]}\,\boldsymbol{e}$. With respect to $\boldsymbol{\alpha}$, we expand (17) as follows

$$\boldsymbol{\alpha}^T \begin{bmatrix} \bar{\boldsymbol{\Psi}}_{11} & \boldsymbol{0} & \boldsymbol{\mathcal{P}}_{[\hat{N}]}\boldsymbol{B}_{[\hat{N}]}\hat{\boldsymbol{L}}_{[m]} & \boldsymbol{\Omega}_{[\hat{N}]}\hat{\boldsymbol{L}}_{[n]} & \boldsymbol{\mathcal{P}}_{[\hat{N}]}\boldsymbol{B}_{[\hat{N}]}\hat{\boldsymbol{L}}_{[m]} \\ * & \boldsymbol{0} & \boldsymbol{0} & \boldsymbol{0} & \boldsymbol{0} \\ * & * & \boldsymbol{0} & \boldsymbol{0} & \boldsymbol{0} \\ * & * & * & \boldsymbol{0} & \boldsymbol{0} \\ * & * & * & * & \boldsymbol{0} \end{bmatrix} \boldsymbol{\alpha} < 0. \tag{18}$$

where $\bar{\boldsymbol{\Psi}}_{11} = \boldsymbol{A}_{[\hat{N}]}^T\boldsymbol{\mathcal{P}}_{[\hat{N}]} + \boldsymbol{\mathcal{P}}_{[\hat{N}]}\boldsymbol{A}_{[\hat{N}]} + 2\zeta\boldsymbol{\mathcal{P}}_{[\hat{N}]} + \boldsymbol{\Omega}_{[\hat{N}]}\mathbb{L}_{[n]} + (\boldsymbol{\Omega}_{[\hat{N}]}\mathbb{L}_{[n]})^T$ and $\hat{\boldsymbol{L}}_{[m]} = \hat{\boldsymbol{L}}\otimes\boldsymbol{I}_m$, with $\boldsymbol{\Omega} = \boldsymbol{\mathcal{P}}\boldsymbol{B}\boldsymbol{K}$ and $\boldsymbol{\Omega}_{[\hat{N}]} = \boldsymbol{I}_{N-1}\otimes\boldsymbol{\Omega}$. Based on definition of $\boldsymbol{\sigma}_j$, $(1 \leq j \leq 3)$, and using Lemma 2, we formulate the following quadratic constraints

$$\boldsymbol{\sigma}_1^T\boldsymbol{\sigma}_1 = \boldsymbol{x}_{[r]}^T\boldsymbol{\mathcal{L}}_{[n]}^T\,\boldsymbol{\Delta}_K^T\,\boldsymbol{\Delta}_K\,\boldsymbol{\mathcal{L}}_{[n]}\boldsymbol{x}_{[r]} \leq \delta^2\,\lambda_N^2\,\boldsymbol{x}_{[r]}^T\boldsymbol{x}_{[r]}, \tag{19}$$

$$\boldsymbol{\sigma}_2^T\boldsymbol{\sigma}_2 = \boldsymbol{e}^T\,\boldsymbol{L}_{[n]}^T\boldsymbol{L}_{[n]}\,\boldsymbol{e} \leq \lambda_N^2\,\boldsymbol{e}^T\,\boldsymbol{e}, \tag{20}$$

$$\boldsymbol{\sigma}_3^T\boldsymbol{\sigma}_3 = \boldsymbol{e}^T\boldsymbol{L}_{[n]}^T\boldsymbol{\Delta}_K^T\boldsymbol{\Delta}_K\boldsymbol{L}_{[n]}\,\boldsymbol{e} \leq \delta^2\lambda_N^2\,\boldsymbol{e}^T\,\boldsymbol{e}. \tag{21}$$

Furthermore, the event-triggering constraint (9) can be re-written in the following quadratic form

$$\boldsymbol{e}^T\,\boldsymbol{e} \leq \Big(\boldsymbol{M}_{[n]}\hat{\boldsymbol{L}}_{[n]}\boldsymbol{\Lambda}(t)\,\hat{\boldsymbol{x}}(t)\Big)^T\phi^2\Big(\boldsymbol{M}_{[n]}\hat{\boldsymbol{L}}_{[n]}\boldsymbol{\Lambda}(t)\,\hat{\boldsymbol{x}}(t)\Big), \quad (22)$$

Using $\boldsymbol{\Lambda}(t)\hat{\boldsymbol{x}}(t)=\boldsymbol{e}(t)+\boldsymbol{x}(t)$, the following inequality is obtained from (22)

$$\begin{aligned}\boldsymbol{e}^T\,\boldsymbol{e} &\leq (\boldsymbol{x}_{[r]}+\hat{\boldsymbol{L}}_{[n]}\boldsymbol{e})^T\boldsymbol{M}_{[n]}^T\phi^2\boldsymbol{M}_{[n]}(\boldsymbol{x}_{[r]}+\hat{\boldsymbol{L}}_{[n]}\boldsymbol{e}\,)\\ &\leq \|\boldsymbol{M}_{[n]}\|^2(\boldsymbol{x}_{[r]}+\hat{\boldsymbol{L}}_{[n]}\boldsymbol{e})^T\phi^2(\boldsymbol{x}_{[r]}+\hat{\boldsymbol{L}}_{[n]}\boldsymbol{e}). \end{aligned} \quad (23)$$

Considering decision variables $\tau_c$, $(1\leq c\leq 4)$, we use the *S-procedure Lemma* [17, Section 2.6.3] to, respectively, include (19), (20), (21), and (23) in (18). The following matrix inequality is then obtained by applying the *Schur complement* Lemma [17, Section 2.1] and pre and post-multiplying the resulting matrix inequality with $\boldsymbol{Y} = \text{diag}\,(\boldsymbol{I},\boldsymbol{I},\boldsymbol{I},\boldsymbol{I},\boldsymbol{I},\tau_4^{-1}\phi^{-1}\boldsymbol{I})$

$$\begin{bmatrix}\hat{\boldsymbol{\Psi}}_{11} & \boldsymbol{0} & \boldsymbol{\mathcal{P}}_{[\hat{N}]}\boldsymbol{B}_{[\hat{N}]}\hat{\boldsymbol{L}}_{[m]} & \boldsymbol{\Omega}_{[\hat{N}]}\hat{\boldsymbol{L}}_{[n]} & \boldsymbol{\mathcal{P}}_{[\hat{N}]}\boldsymbol{B}_{[\hat{N}]}\hat{\boldsymbol{L}}_{[m]} & \|\boldsymbol{M}_{[n]}\|\boldsymbol{I}\\ * & \hat{\boldsymbol{\Psi}}_{22} & \boldsymbol{0} & \boldsymbol{0} & \boldsymbol{0} & \|\boldsymbol{M}_{[n]}\|\hat{\boldsymbol{L}}_{[n]}^T\\ * & * & -\tau_1\boldsymbol{I} & \boldsymbol{0} & \boldsymbol{0} & \boldsymbol{0}\\ * & * & * & -\tau_2\boldsymbol{I} & \boldsymbol{0} & \boldsymbol{0}\\ * & * & * & * & -\tau_3\boldsymbol{I} & \boldsymbol{0}\\ * & * & * & * & * & -\tau_4^{-1}\phi^{-2}\boldsymbol{I}\end{bmatrix}<0, \quad (24)$$

where $\hat{\boldsymbol{\Psi}}_{11} = \bar{\boldsymbol{\Psi}}_{11} + \tau_1\lambda_N^2\delta^2\boldsymbol{I}$, and $\hat{\boldsymbol{\Psi}}_{22} = (\tau_2\lambda_N^2 + \tau_3\lambda_N^2\delta^2 - \tau_4)\boldsymbol{I}$. Due to product of decision variables in $\tau_4^{-1}\phi^{-2}$, the matrix inequality (24) is not affine. Using $\gamma\triangleq\tau_4^{-1}\phi^{-2}$ as an alternative variable, inequality (24) becomes affine with respect to $\gamma$. The matrix inequality (24) is the sufficient constraint to guarantee resilient $\zeta$-exponential event-triggered consensus for (7). However, global information is required to compute $\boldsymbol{L}$ and its variants. Using Lemma 2 and under condition $\boldsymbol{\Omega}<0$, it holds that $\boldsymbol{\Omega}_{[\hat{N}]}\mathbb{L}_{[n]}+(\boldsymbol{\Omega}_{[\hat{N}]}\mathbb{L}_{[n]})^T\leq 2\lambda_2\boldsymbol{\Omega}\boldsymbol{I}$ Next, we alter matrix inequality (24) to the following structure

$$\tilde{\boldsymbol{\Psi}} + \sum_{j=1}^{3}\boldsymbol{\mathcal{S}}_j^T\boldsymbol{\mathcal{R}}_j + \sum_{j=1}^{3}\boldsymbol{\mathcal{R}}_j^T\boldsymbol{\mathcal{S}}_j < 0, \quad (25)$$

where

$$\tilde{\boldsymbol{\Psi}} = \begin{bmatrix}\tilde{\boldsymbol{\Psi}}_{11} & \boldsymbol{0} & \boldsymbol{0} & \boldsymbol{0} & \boldsymbol{0} & \|\boldsymbol{M}_{[n]}\|\boldsymbol{I}\\ * & \hat{\boldsymbol{\Psi}}_{22} & \boldsymbol{0} & \boldsymbol{0} & \boldsymbol{0} & \boldsymbol{0}\\ * & * & -\tau_1\boldsymbol{I} & \boldsymbol{0} & \boldsymbol{0} & \boldsymbol{0}\\ * & * & * & -\tau_2\boldsymbol{I} & \boldsymbol{0} & \boldsymbol{0}\\ * & * & * & * & -\tau_3\boldsymbol{I} & \boldsymbol{0}\\ * & * & * & * & * & -\gamma\boldsymbol{I}\end{bmatrix}, \quad (26)$$

with $\tilde{\boldsymbol{\Psi}}_{11}=\boldsymbol{A}_{[\hat{N}]}^T\boldsymbol{\mathcal{P}}_{[\hat{N}]}+\boldsymbol{\mathcal{P}}_{[\hat{N}]}\boldsymbol{A}_{[\hat{N}]}+2\zeta\boldsymbol{\mathcal{P}}_{[\hat{N}]}+2\lambda_2\boldsymbol{\Omega}\boldsymbol{I}+\tau_1\lambda_N^2\delta^2\boldsymbol{I}$, and $\boldsymbol{\mathcal{S}}_1=[\,\boldsymbol{0},\boldsymbol{0},\boldsymbol{\mathcal{P}}_{[\hat{N}]},\boldsymbol{0},\boldsymbol{\mathcal{P}}_{[\hat{N}]},\boldsymbol{0}\,]$, $\boldsymbol{\mathcal{R}}_1=[\,\boldsymbol{B}_{[\hat{N}]}\hat{\boldsymbol{L}}_{[m]},\boldsymbol{0},\boldsymbol{0},\boldsymbol{0},\boldsymbol{0},\boldsymbol{0}\,]$, $\boldsymbol{\mathcal{S}}_2=[\,\boldsymbol{\Omega}\boldsymbol{I},\,\boldsymbol{0},\,\boldsymbol{0},\,\boldsymbol{0},\,\boldsymbol{0},\,\boldsymbol{0}\,]$, $\boldsymbol{\mathcal{R}}_2=[\,\boldsymbol{0},\,\boldsymbol{0},\hat{\boldsymbol{L}}_{[n]},\,\boldsymbol{0},\,\boldsymbol{0},\,\boldsymbol{0}\,]$, $\boldsymbol{\mathcal{S}}_3=[\,\boldsymbol{0},\boldsymbol{0},\boldsymbol{0},\boldsymbol{0},\boldsymbol{0},\|\boldsymbol{M}_{[n]}\|\boldsymbol{I}\,]$, and $\boldsymbol{\mathcal{R}}_3=[\,\boldsymbol{0},\,\hat{\boldsymbol{L}}_{[n]},\boldsymbol{0},\boldsymbol{0},\boldsymbol{0},\boldsymbol{0}\,]$. According to [18, Lemma 2.1], inequality (25) holds iff there exist positive scalars $\epsilon_j$, $(1\leq j\leq 3)$, such that

$$\tilde{\boldsymbol{\Psi}} + \sum_{j=1}^{3}\epsilon_j^{-1}\boldsymbol{\mathcal{S}}_j^T\boldsymbol{\mathcal{S}}_j + \sum_{j=1}^{3}\epsilon_j\,\boldsymbol{\mathcal{R}}_j^T\boldsymbol{\mathcal{R}}_j < 0. \quad (27)$$

Let $\bar{\boldsymbol{\mathcal{R}}}_1=[\,\lambda_N\boldsymbol{B},\boldsymbol{0},\boldsymbol{0},\boldsymbol{0},\boldsymbol{0},\boldsymbol{0}\,]$ $\bar{\boldsymbol{\mathcal{R}}}_2=[\,\boldsymbol{0},\boldsymbol{0},\lambda_N\boldsymbol{I},\,\boldsymbol{0},\,\boldsymbol{0},\,\boldsymbol{0}\,]$, and $\bar{\boldsymbol{\mathcal{R}}}_3=[\,\boldsymbol{0},\lambda_N\boldsymbol{I},\boldsymbol{0},\boldsymbol{0},\boldsymbol{0},\boldsymbol{0}\,]$. Then, it holds that $\boldsymbol{\mathcal{R}}_j^T\boldsymbol{\mathcal{R}}_j\leq\bar{\boldsymbol{\mathcal{R}}}_j^T\bar{\boldsymbol{\mathcal{R}}}_j$, $(1\leq j\leq 3)$. Therefore, if the following inequality is guaranteed, then (27) is also guaranteed

$$\tilde{\boldsymbol{\Psi}} + \sum_{j=1}^{3}\epsilon_j^{-1}\boldsymbol{\mathcal{S}}_j^T\boldsymbol{\mathcal{S}}_j + \sum_{j=1}^{3}\epsilon_j\,\bar{\boldsymbol{\mathcal{R}}}_j^T\bar{\boldsymbol{\mathcal{R}}}_j < 0. \quad (28)$$

Three times using the *Schur complement* Lemma for (28) results in $\boldsymbol{\Psi}<0$, where $\boldsymbol{\Psi}$ is defined in (13). As stated previously, an ideal objective function for $\boldsymbol{\Psi}<0$ would attempt to maximize $\phi$ and minimize norm of $\boldsymbol{K}$. In addition to the multi-objective nature of the problem, the change of variables used to derive $\boldsymbol{\Psi}$ makes such an objective function nonlinear. Motivated by the proposed approaches in [19, Section 3] and [20, Section 2.2], a linear scalarization method is used to minimize the decision variables involved in obtaining $\boldsymbol{K}$ and $\phi$. To this end, the following constraints are considered for $\tau_4$, $\gamma$, $\boldsymbol{\mathcal{P}}$, and $\boldsymbol{\Omega}$

$$\begin{aligned}&\{\,\boldsymbol{\mathcal{P}}^{-1}<\theta_1\boldsymbol{I},\quad \theta_1>0\,\},\quad \{\,\boldsymbol{\Omega}^T\boldsymbol{\Omega}<\theta_2\boldsymbol{I},\quad \theta_2>0\,\},\\ &\{\,\tau_4<\theta_3,\quad \theta_3>0\,\},\quad \{\,\gamma<\theta_4,\quad \theta_4>0\,\},\end{aligned} \quad (29)$$

where $\theta_c$, $(1\leq c\leq 4)$, is a decision variable. Considering (29), the proposed objective function $\mathcal{F}$ in (12) minimizes a linear combination of weights for bounding parameters $\theta_c$. All weights are set to one to treat $\boldsymbol{K}$ and $\phi$ equally. It is known that the solution of (12) is a Pareto optimal point for $\mathcal{F}$ [20]. The *Schur complement* is used to convert the matrix inequalities in (29) into the LMI structures given in (14). Once (12) is solved, consensus parameters are computed from (11). □

Based on Theorem 1, the proposed event-triggered consensus approach is summarized in Algorithm 1.

---

**Algorithm 1 . Performance Constrained Distributed Event-trigged Consensus.**

---

**I. Parameter Design: (D1 -D3)**

D1. *Preliminaries*: Use an average consensus [2] to agree on $\zeta$. Use distributed approaches such as [21]–[23] to locally estimate $\delta$, $N$, $\lambda_2$, and $\lambda_N$.

D2. *Optimization*: Using a semidefinite programming (SDP) solver, solve (12) for given $\zeta$ and $\delta$.

D3. *Feasibility Verification*: If a solution exists for (12), compute $\phi$ and $\boldsymbol{K}$ from (11). Otherwise, reduce $\zeta$ and/ or $\delta$. Then, repeat step D2.

**II. Execution: (E1 and E2)**

E1. *Initialization*: All agents transmit their initial state values $\boldsymbol{x}_i(0)$ to the neighbours.

E2. *Consensus process*: Using designed $\boldsymbol{K}$ for (4) and $\phi$ for (3) the states of the agents approach a consensus.

---

**Remark 1.** To solve optimization (12) locally, each node should recognize $\zeta$, $\lambda_2$, $\lambda_N$, $N$, and $\delta$. Parameter $\zeta$ can be agreed among the agents using a distributed average-consensus algorithm [2]. We note that eigenvalues $\lambda_2$ and $\lambda_N$ can be estimated with arbitrary sufficient accuracy in a distributed fashion using the proposed approach in [21]. A distributed algorithm is developed in [22] to accurately estimate the network size $N$. Moreover, the value of $\delta$ can be computed distributively using a max-consensus algorithm [23]. As a result, optimization (12) is performed locally followed by these preliminary steps.

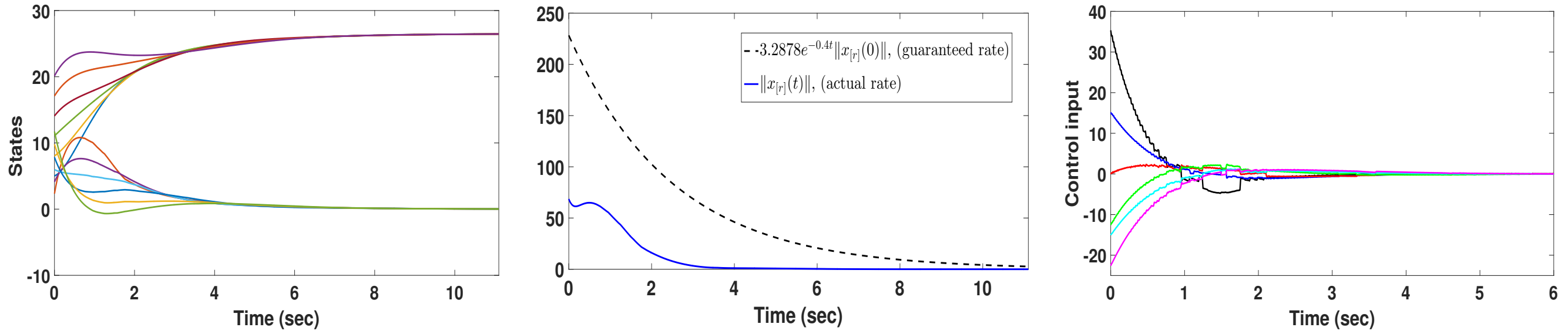

Fig. 1: Trajectories of the system; (a): Position $r_i(t)$ and Velocity $v_i(t)$, (b): Exponential consensus, (C): Control input $u_i(t)$.

**Remark 2.** In event-triggered schemes there must be a finite number of events within a finite time interval. Otherwise, the event-triggering mechanism exhibits Zeno behavior [15]. It is proved in [14, Theorem 2] that the proposed scheme does not exhibit Zeno behaviour. According to [14, Eq. (29)], the events are more sparse for larger values of $\phi$ and smaller norms of $\boldsymbol{K}$. This fact is consistent with the proposed objective function $\mathcal{F}$, which enlarges the value of $\phi$ and restricts the norm of $\boldsymbol{K}$. Therefore, $\mathcal{F}$ helps reducing the number of events by enlarging the minimum inter-event interval.

**Remark 3.** Unlike our previous work [14], where complexity of optimization grows with the network size $N$, complexity of solving (12) does not depend on $N$. Beside considerably reducing the processing time to solve (12), this accomplishment improves scalability of the approach, especially for large networks. More precisely, the computational effort for solving a semidefinite programming (SDP) problem (such as (12)) via interior-point methods depends on (a) Total number of iterations required to approach the optimal solution, and; (b) The order of arithmetic operations performed in each iteration. Let parameter $d_m$ define the "highest dimension among the LMI constraints in an SDP problem", e.g., for (12) $d_m = 9n$ (dimension of matrix $\boldsymbol{\Psi}$). Also, let $n_v$ denote the "total number of decision variables used in an SDP problem". For (12), $n_v = n(n+1)+12$. According to [24], in the worst-case complexity the number of iterations required to approach the optimal solution grows at the $O(\sqrt{N_p}\,|\log \epsilon_g|)$ where $N_p = \max\{d_m, n_v\}$, and $\epsilon_g$ is the duality gap which refers to the difference between the value of optimal solutions for primal and dual problems. This parameter (i.e., $\epsilon_g$) is often used as an indication of accuracy in solving SDP problems [24]. As for the arithmetic operation needed in each iteration, the structure of matrix constraints plays a significant role. As a rule of thumb, each iteration requires on the $O(\max\{d_m^3, d_m^2 n_v, F\})$ operations, where $F$ is the cost of evaluating the first and second derivatives of the constraints and objective function [25, Section 1.3]. Since none of the above parameters depend on $N$, the complexity of (12) does not grow with $N$.

## IV. Simulation

Consider a network of agents with dynamics

$$\left.\begin{aligned} \dot{r}_i(t) &= v_i(t), \\ \dot{v}_i(t) &= -\mu_k v_i(t) + u_i(t). \end{aligned}\right\} \quad 1 \le i \le 6 \tag{30}$$

where $r_i(t) \in \mathbb{R}$ and $v_i(t) \in \mathbb{R}$, respectively, denote the position and velocity of agent $i$. Scalar $\mu_k > 0$ denotes the velocity damping and $u_i(t)$ is the control input. Considering $\mu_k = 0.5$, the state space representation for (30) is given by $\boldsymbol{x}_i(t) = [r_i(t), v_i(t)]^T$, $\boldsymbol{A} = [\,0, 1;\ 0, -0.5\,]$, and $\boldsymbol{B} = [\,0, 1\,]^T$. Let $\boldsymbol{L} = [\,4, 0, -1, -1, -1, -1;\ 0, 2, 0, -1, 0, -1;\ -1, 0, 2, 0, -1, 0;\ -1, -1, 0, 2, 0, 0;\ -1, 0, -1, 0, 2, 0;\ -1, -1, 0, 0, 0, 2\,]$. To solve consensus using Algorithm 1, we initialize optimization (12) with $\zeta = 0.4$ and $\delta = 0.002$. It can be verified that $\lambda_2 = 0.7639$ and $\lambda_N = 5.2361$. For an accuracy of $\epsilon_g = 10^{-8}$, it takes 17 iterations (with a total time of 0.7411 seconds) for the SDPT3 solver [26] to solve (12) with the following solution

$$\begin{aligned} &\boldsymbol{\mathcal{P}} = \begin{bmatrix} 0.0278 & -0.0265 \\ -0.0265 & 0.2657 \end{bmatrix}, \boldsymbol{\Omega} = \begin{bmatrix} -0.1283 & -0.0083 \\ -0.0083 & -0.0990 \end{bmatrix}, \tau_1 = 34.33, \\ &\tau_2 = 4.68,\ \tau_3 = 3810.85,\ \tau_4 = 141.6719,\ \epsilon_1 = 0.00213, \\ &\epsilon_2 = 0.1708,\ \epsilon_3 = 0.4678,\ \gamma = 100.7461,\ \theta_1 = 40.2484, \\ &\theta_2 = 0.01703,\ \theta_3 = 141.6719,\ \theta_4 = 100.7461, \end{aligned} \tag{31}$$

Using (11) and solution (31), we calculate $\boldsymbol{K} = -[0.5446,\ 0.4450]$ and $\phi = 0.0084$. Moreover, $\eta = 3.2878$ and $\mathcal{F} = 282.68$. Let $\boldsymbol{x}_i(0) = [3i{+}2,\ 2i]^T$, $(1 \le i \le 6)$. Control gain uncertainty is assumed is sinusoidal form satisfying $\|\boldsymbol{\Delta}_K\| \le \delta$. In Fig.1(a), we plot trajectories of (30) reaching consensus. Fig.1(b) is included to verify that the obtained parameters are capable of ensuring 0.4-exponential consensus. In Fig.1(c), the control inputs used to achieve consensus are plotted. For a termination threshold $\|\boldsymbol{x}_{[r]}(t)\| \le 0.01$, it takes 11.112 sec to reach consensus. Let $\overline{\mathrm{CT}}$ denote the time at which consensus is reached, e.g., $\overline{\mathrm{CT}} = 11.112$ in this example. Simulation results show that the six agents, respectively, trigger 337, 344, 401, 402, 349, and 358, events during the process. The average number of events is $\overline{\mathrm{AE}} = 365.17$. We also define $\overline{\mathrm{AET}} = \overline{\mathrm{AE}}/\overline{\mathrm{CT}}$, which refers to the average number of evens triggered per unit time. In this example, $\overline{\mathrm{AET}} = 32.86$. Since in practice the actuators are often limited to operate within a restricted region, we report the maximum control input used by the agents calculated by $\bar{\boldsymbol{u}} = \max_{1 \le i \le 6,\, t \ge 0} |\boldsymbol{u}_i(t)|$. In this case, $\bar{\boldsymbol{u}} = 35.43$.

**Effect of $\zeta$ and** $\delta$: Next, we study the effect of different values for $\zeta$ and $\delta$ on consensus features. To this end, we solve optimization (12) for given values of $\zeta$ and $\delta$ in Table II. Then, we run consensus simulation for (30) with corresponding computed parameters. The following facts are observed according to Table II: (i) As expected, the value for $\overline{\mathrm{CT}}$ is decreased as a larger convergence rate is selected. (ii) The values $\overline{\mathrm{AE}}$ and $\overline{\mathrm{AET}}$ are relatively increased when a larger convergence rate is selected. (iii) As $\zeta$ is increased, $\mathcal{F}$ becomes larger. This implies computing a larger $\boldsymbol{K}$ and/or a smaller $\phi$. As a result, the value for $\bar{\boldsymbol{u}}$ and $\overline{\mathrm{AET}}$ is increased. (iv) A larger value for $\delta$ leads to a larger $\mathcal{F}$. Intuitively speaking, larger control gains are needed to compensate the effect of a larger uncertainty in their nominal values. As a result, the control input trajectories are more fluctuating and $\bar{\boldsymbol{u}}$ is increased; (v) The consensus time $\overline{\mathrm{CT}}$ is decreased as $\delta$ is increased. The reason lies in that larger control gains provide larger control inputs and a faster consensus. These results verify the flexibility and effectiveness of the proposed performance guaranteed approach in maintaining the event-triggered consensus.

TABLE II: Consensus performance for different $\zeta$ and $\delta$.

| $\zeta$ | $\delta$ | $\overline{\mathrm{CT}}$ | $\overline{\mathrm{AE}}$ | $\overline{\mathrm{AET}}$ | $\bar{\boldsymbol{u}}$ | $\mathcal{F}$ |
|---|---|---|---|---|---|---|
| 0.2 | 0.002 | 27.533 | 115.83 | 4.22 | 8.73 | 264.18 |
| 0.3 | 0.002 | 15.039 | 311.50 | 20.71 | 21.40 | 274.29 |
| 0.4 | 0.002 | 11.112 | 365.17 | 32.86 | 35.43 | 282.68 |
| 0.5 | 0.002 | 9.563 | 431.83 | 45.21 | 50.50 | 290.00 |
| 0.4 | 0.001 | 18.497 | 352.17 | 19.02 | 16.23 | 276.25 |
| 0.4 | 0.002 | 11.112 | 365.17 | 32.86 | 35.43 | 282.68 |
| 0.4 | 0.003 | 9.017 | 223.17 | 24.72 | 57.36 | 287.15 |

## V. Conclusion

This paper proposes a distributed event-triggered consensus approach for linear multi-agent systems that guarantees a desired exponential rate of consensus and some resilience to control gain perturbation. Due to mutual effect of the control law and event-triggering scheme on consensus performance, design parameters, namely the event-triggering threshold (ET) and control gain, are obtained through a *unified* constrained convex optimization framework with a proposed objective function incorporating both terms in a linear weighted combination. The optimization is independent of the knowledge of network Laplacian matrix and can be performed distributively. Moreover, complexity of the proposed optimization does not grow with network size. Simulation results demonstrate flexibility of the proposed approach in reaching event-triggered consensus with performance constraints.

## References

[1] F. Garin and L. Schenato, "A survey on distributed estimation and control applications using linear consensus algorithms," *Networked control systems.* Springer, 2010, pp. 75–107.

[2] M.M. Asadi, S. Blouin and A.G. Aghdam, "Distributed Dynamic Average Consensus in Asymmetric Networks," *Annual American Control Conference (ACC)*, 2018, pp. 864-869.

[3] S. Rao and D. Ghose, "Sliding mode control-based autopilots for leaderless consensus of unmanned aerial vehicles," *IEEE Trans. Control Syst. Technol.*, vol. 22, pp. 1964-1972, 2014.

[4] S. Chen and D. W. Ho, "Consensus control for multiple auvs under imperfect information caused by communication faults," *Inf. Sci.*, vol. 370, pp. 565–577, 2016.

[5] L. Ding, Q.-L. Han, X. Ge, X.-M. Zhang, "An overview of recent advances in event-triggered consensus of multiagent systems," *IEEE Trans. Cybern.*, vol. 48, pp. 1110-1123, 2018.

[6] D. Liu and G.-H. Yang, "Event-triggered non-fragile control for linear systems with actuator saturation and disturbances," *Inf. Sci.*, vol. 429, pp. 1–11, 2018.

[7] H. Shen, F. Li, Z.-G. Wu, J. H. Park, and V. Sreeram, "Fuzzy-model-based non-fragile control for nonlinear singularly perturbed systems with semi-markov jump parameters," *IEEE Trans. Fuzzy Systems*, 2018.

[8] H. Li, G. Chen, Z. Dong, and D. Xia, "Consensus analysis of multiagent systems with second-order nonlinear dynamics and general directed topology: An event-triggered scheme," *Inf. Sci.*, vol. 370, pp. 598–622, 2016.

[9] H. Yu and F. Hao, "Design of event conditions in event-triggered control systems: a non-fragile control system approach," *IET Control Theory & Applications*, vol. 10, no. 9, pp. 1069–1077, 2016.

[10] W. Hu, L. Liu, and G. Feng, "Consensus of linear multi-agent systems by distributed event-triggered strategy," *IEEE Trans. Cybern.*, vol. 46, no. 1, pp. 148–157, 2016.

[11] P. Yu, L. Ding, Z.-W. Liu, and Z.-H. Guan, "A distributed event-triggered transmission strategy for exponential consensus of general linear multi-agent systems with directed topology," *J. Franklin I.*, vol. 352, no. 12, pp. 5866–5881, 2015.

[12] B. Zhou, X. Liao, T. Huang, and G. Chen, "Leader-following exponential consensus of general linear multi-agent systems via event-triggered control with combinational measurements," *Appl. Math. Lett.*, vol. 40, pp. 35–39, 2015.

[13] L. Ding, Q.-L. Han, L. Y. Wang, and E. Sindi, "Distributed cooperative optimal control of dc microgrids with communication delays," *IEEE Trans. Ind. Informat.*, 2018.

[14] A. Amini, A. Mohammadi, and A. Asif, "Resilient event-triggered consensus with exponential convergence in multi-agent systems," in *IEE Annual American Control Conference (ACC)*, 2018, pp. 2889-2896.

[15] D. Yang, W. Ren, X. Liu, W. Chen, "Decentralized event-triggered consensus for linear multi-agent systems under general directed graphs," *Automatica*, vol. 69, pp. 242-249, 2016.

[16] V. Phat, Y. Khongtham, and K. Ratchagit, "Lmi approach to exponential stability of linear systems with interval time-varying delays," *Linear Algebra Appl.*, vol. 436, no. 1, pp. 243–251, 2012.

[17] S. Boyd, L. El Ghaoui, E. Feron, and V. Balakrishnan, *Linear matrix inequalities in system and control theory.* SIAM, 1994.

[18] C.H. Lien, "Non-fragile guaranteed cost control for uncertain neutral dynamic systems with time-varying delays in state & control input," *Chaos Sol. Fract.*, vol. 31, pp. 889-899, 2007.

[19] D. Šiljak and D. Stipanovic, "Robust stabilization of nonlinear systems: the lmi approach," *Mathematical problems in Engineering*, vol. 6, no. 5, pp. 461–493, 2000.

[20] G. Chiandussi, M. Codegone, S. Ferrero, and F. E. Varesio, "Comparison of multi-objective optimization methodologies for engineering applications," *Computers & Mathematics with Applications*, vol. 63, no. 5, pp. 912–942, 2012.

[21] A. Gusrialdi and Z. Qu, "Distributed estimation of all the eigenvalues and eigenvectors of matrices associated with strongly connected digraphs," *IEEE control systems letters*, vol. 1, no. 2, pp. 328–333, 2017.

[22] I. Shames, T. Charalambous, C. N. Hadjicostis, and M. Johansson, "Distributed network size estimation and average degree estimation and control in networks isomorphic to directed graphs," in *Communication, Control, and Computing (Allerton).* IEEE, 2012, pp. 1885–1892.

[23] S. Zhang, C. Tepedelenlioglu, M. K. Banavar, A. Spanias, "Max-consensus using the soft maximum," *IEEE Asilomar Conf. Signals, Systems & Computers*, pp. 433-437.

[24] F. Alizadeh, "Interior point methods in semidefinite programming with applications to combinatorial optimization," *SIAM journal on Optimization*, vol. 5, no. 1, pp. 13–51, 1995.

[25] S. Boyd and L. Vandenberghe, *Convex optimization.* Cambridge university press, 2004.

[26] K.-C. Toh, M. J. Todd, and R. H. Tütüncü, "Sdpt3-a matlab software package for semidefinite programming, version 1.3," *Optimization Methods & Software*, vol. 11, pp. 545-581, 1999.